\documentclass[aps,twocolumn,longbibliography]{revtex4-1}
\usepackage{graphicx}
\usepackage{amsmath,amssymb}
\usepackage{dsfont}
\usepackage{bm,lipsum}
\usepackage{color,xcolor}
\usepackage{physics}
\usepackage{MnSymbol, marvosym, wasysym}
\usepackage{hyperref}
\usepackage[caption=false]{subfig}
\usepackage{float,color}
\usepackage{soul}

\AtBeginDocument{%
    \newwrite\bibnotes
    \def\bibnotesext{Notes.bib}
    \immediate\openout\bibnotes=\jobname\bibnotesext
    \immediate\write\bibnotes{@CONTROL{REVTEX41Control}}
    \immediate\write\bibnotes{@CONTROL{%
    apsrev41Control,author="08",editor="1",pages="1",title="0",year="1"}}
     \if@filesw
     \immediate\write\@auxout{\string\citation{apsrev41Control}}%
    \fi
}%

\begin{document}

\renewcommand{\vec}[1]{\boldsymbol{#1}}
\newcommand{\up}{{\uparrow}}
\newcommand{\dw}{{\downarrow}}
\newcommand{\pa}{{\partial}}
\newcommand{\pd}{{\phantom{\dagger}}}
\newcommand{\bs}[1]{\boldsymbol{#1}}
\newcommand{\todo}[1]{{\textbf{\color{red}ToDo: #1}}}
\newcommand{\new}[1]{{\textbf{\color{red}#1}}}
\newcommand{\sr}[1]{{\color{blue}#1}}
\newcommand{\srr}[1]{{\color{orange}#1}}
\newcommand{\eps}{{\varepsilon}}
\newcommand{\I}{{i\mkern3mu}}
\newcommand{\nn}{\nonumber}
\newcommand{\ie}{{\it i.e.},\ }
\def\eg{\emph{e.g.}\ }
\def\ea{\emph{et al.}}
\def\cf{\emph{c.f.}\ }


\title{Realization of a discrete time crystal on 57 qubits of a quantum computer}

\author{Philipp Frey}
\author{Stephan Rachel}
\affiliation{School of Physics, University of Melbourne, Parkville, VIC 3010, Australia}

\date{\today}


\maketitle


\textbf{
Novel dynamical phases that violate ergodicity have been a subject of extensive research in recent years. A periodically driven system is naively expected to lose all memory of its initial state due to thermalization, yet this can be avoided in the presence of many-body localization. A discrete time crystal represents a driven system whose local observables spontaneously break time translation symmetry and retain memory of the initial state indefinitely.
Here we report the observation of a discrete time crystal on a chain consisting of 57 superconducting qubits on a state--of--the--art quantum computer. We probe random initial states
and compare the cases of vanishing and finite disorder to distinguish many-body localization from pre-thermal dynamics. We further report results on the dynamical phase transition between the discrete time crystal and a thermal regime, which is observed via critical fluctuations in the system's sub-harmonic frequency response and a significant speed-up of spin depolarisation.}

The phenomenon of spontaneous symmetry breaking is common in nature and characterizes a large class of phases
in materials. For instance, the crystal lattice of a solid formed by the nuclei breaks the continuous spatial translation symmetry of the underlying Hamiltonian, as indicated by the mere discrete translation symmetry of the local density operator. In 2012, Wilczek proposed the existence of phases of matter that break continuous time translation symmetry\,\cite{Wilczek_2012}, dubbed time crystals. Later their existence in thermal equilibrium was ruled out\,\cite{PhysRevLett.114.251603}. However, time crystals can be stabilized as out-of-equilibrium matter such as periodically driven systems\,\cite{khemani2019brief, Sacha_2017,PhysRevLett.118.030401}, where the dynamical symmetry corresponds to discrete time translations only. Such discrete time crystals (DTC)\,\cite{khemani-16prl250401, PhysRevLett.117.090402,Moessner2017, Choi2017, else-20arcmp467} 
exhibit period doubling, tripling etc.\ with respect to the periodic driving Hamiltonian. That is, the system returns to its initial state only after some integer multiple of the driving period. Ergodicity predicts that  driving heats the system, causing it to thermalize after a certain number of periods. Instead, a DTC is not ergodic and ``remembers'' its initial state even at late times, in violation of the eigenstate thermalization hypothesis\,\cite{srednicki-94pre888}.

In order to escape ergodicity, a DTC must be many-body localized (MBL). MBL has been a subject of extensive research in recent years, both theoretical and experimental\,\cite{
PhysRevB.75.155111,  
PhysRevB.76.052203,  
PhysRevB.82.174411,  
PhysRevLett.95.206603, 
PhysRevB.98.174202, 
 PhysRevB.88.014206,
 Smith2016}.
It can be understood as an emergent integrability through localized integrals of motion. In their presence, the system is prevented from heating to a state that locally resembles thermal equilibrium. For comprehensive reviews on MBL see
\cite{doi:10.1146/annurev-conmatphys-031214-014726, 
RevModPhys.91.021001, 
 ALET2018498, 
doi:10.1146/annurev-conmatphys-031214-014701, 
ABANIN2021168415}.

Early and previous experiments have provided important insights on DTCs in a variety of different platforms such as trapped ions\,\cite{Smith2016, Zhang2017,kyprianidis-21s1192}, dipolar spin systems\,\cite{Choi2017,PhysRevLett.120.180603,pal-18prl180602,osullivan-20njp085001,randall-21arXiv} and superfluid quantum gases\,\cite{smits-18prl185301,autti-18prl215301,giergiel-18pra013613}. Most of these experiments fail, however, to satisfy all experimental requirements for realizing DTC spatiotemporal order\,\cite{ippoliti2021manybody}: the systems need to be truly many-body and coherence times must be sufficiently long to be able to distinguish a  DTC from short-time transients; the implemented Hamiltonian must contain disordered spin-spin couplings and sufficiently short-ranged interactions. Firmly establishing DTC dynamics requires the ability to prepare arbitrary initial states and perform site-resolved measurements. Present day quantum computers, so-called noisy intermediate-scale quantum computing (NISQ) devices, have been suggested as the only platform that currently meets all of the requirements above\,\cite{ippoliti2021manybody}. Comparatively short coherence times due to noise in the system pose the predominant challenge.

Here, we report the observation of a DTC over a 57-qubit chain on IBM's quantum computers {\it ibmq\_manhattan} and {\it ibmq\_brooklyn}. The best-studied instance is the one-dimensional spin-1/2 chain with disordered nearest neighbour Ising-interactions, driven by an imperfect periodic spin flip\,\cite{Zhang2017, PhysRevLett.118.030401, PhysRevB.94.085112}.
Prethermal dynamics\,\cite{abanin-17cmp809,kyprianidis-21s1192} can mimic the DTC phenomenon for initial states that lie at edge of the many body spectrum, whereas true MBL applies to the entire spectrum. Therefore it is essential to probe random bit strings as initial states.
By preparing both fully polarized and random-bit states, and varying the simulated disorder, we are able to distinguish between prethermal and the long-sought DTC regime.
We further present results on the dynamical phase transition between MBL-DTC and thermal phase. 

\section*{DTC on a quantum computer}

We implement a peridically driven Ising chain with quenched disorder and imperfect drive. This is an instance of Floquet-evolution, \ie time evolution defined in terms of a unitary $U$ instead of a Hamiltonian $H$.
We make use of the ability to directly program any unitary operator acting on a set of qubits which, due to their nearest neighbor connectivity, is equivalent to a one-dimensional spin chain. The periodic driving can also be thought of as time evolution under a piecewise defined Hamiltonian, resulting in discrete time translation symmetry or periodicity. As we will show, the state of this system itself spontaneously breaks this symmetry through period doubling and therefore represents a DTC.
The time evolution operator $U$ of the Floquet-system is defined in terms of two unitaries $U=U_\mathrm{2}U_\mathrm{1}$:
One represents an imperfect global spin flip,
\begin{equation}\label{eq1}
U_\mathrm{1} = \exp(\I \frac{\pi}{2} (1 - \epsilon) \sum_{i} X_{i})\ ,
\end{equation}
where $X_{i}$ is the Pauli X-gate on the $i$th qubit. The parameter $\epsilon$ accounts for a deviation from an ideal spin flip, since qualitatively novel behavior needs to be robust against small perturbations in order to be considered a true phase of the system. The other unitary corresponds to nearest neighbor Ising interactions,
\begin{equation}\label{eq2}
U_\mathrm{2} = \exp(-\I   \sum_{i} J_{i} Z_{i} Z_{i+1}  )\ ,
\end{equation}
with the  Ising couplings $J_i$ containing the quenched local disorder. The second unitary can be varied by adjusting the Ising interaction couplings $\{J_i\}$. We generate disordered sets of couplings by picking each one randomly from an interval $J_i\in[\pi/8, 3\pi/8]$ centered around the mean $\pi/4$.
While the model defined by the unitaries Eqs.\,\ref{eq1} and \ref{eq2} is integrable and non-interacting, the finite gate errors on the quantum computer introduce effective terms such as longitudinal fields $\sim\exp(i\sum_i b_i Z_i)$; these additional terms make the model non-integrable and truly many-body (see Discussion).

\begin{figure}[t!]
\centering
\includegraphics[width=1\columnwidth]{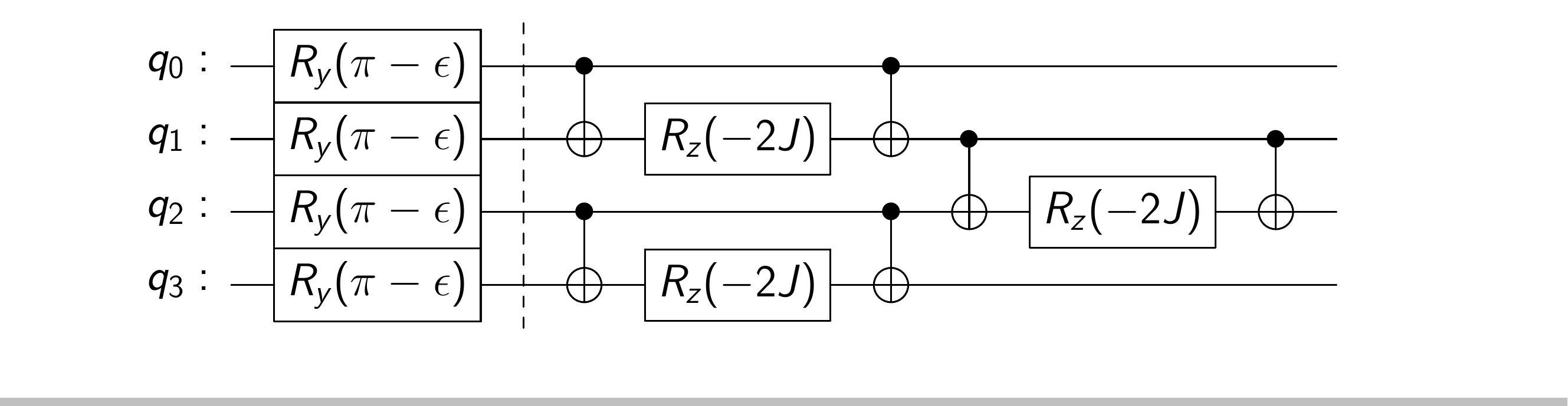}
\caption{\textbf{Illustrative 4-qubit circuit for one Floquet-period.} $R_y$ and $R_z$ represent single-qubit roations around the $y$ and $z$ axis, respectively. Vertical lines connecting small and large open circles represent CNOT gates.  The first part to the left of the vertical dotted line implements the imperfect spin flip while the latter part to the right of this line implements the nearest neighbor Ising interactions.}
\label{fig:circuit floquet}
\end{figure}

A different random bit string as the initial state and a different disorder realization is used for each run of the time evolution. We use a trotterized time evolution, which, for this model, is an exact representation of the unitary $U$. In Fig.\,\ref{fig:circuit floquet} we show the circuit decomposition in terms of basis gates for a Trotter step on four qubits. We then measure each qubit in the computational basis corresponding to $Z$. The error rates on current NISQ devices limits the number of Floquet periods to $\sim 50$. By making use of the heavy hex topology of {\it ibmq\_brooklyn} and {\it ibmq\_manhattan} (see Fig.\,\ref{fig:layout}) we are able to simulate an $N=57$ site chain and to thereby go far beyond current numerical calculations on any classical computer. These machines have stated average CNOT error rates between $1.1\mathrm{e}{-2}$ and $3\mathrm{e}{-2}$
and average readout error rates between $2.5\mathrm{e}{-2}$ and $3.7\mathrm{e}{-2}$.
The circuit for each time step is run 32768 times in order to minimize shot noise.

\vspace{20pt}
\begin{figure*}[t]
\centering
\hspace{13mm}
\subfloat{\includegraphics[width=0.89\linewidth]{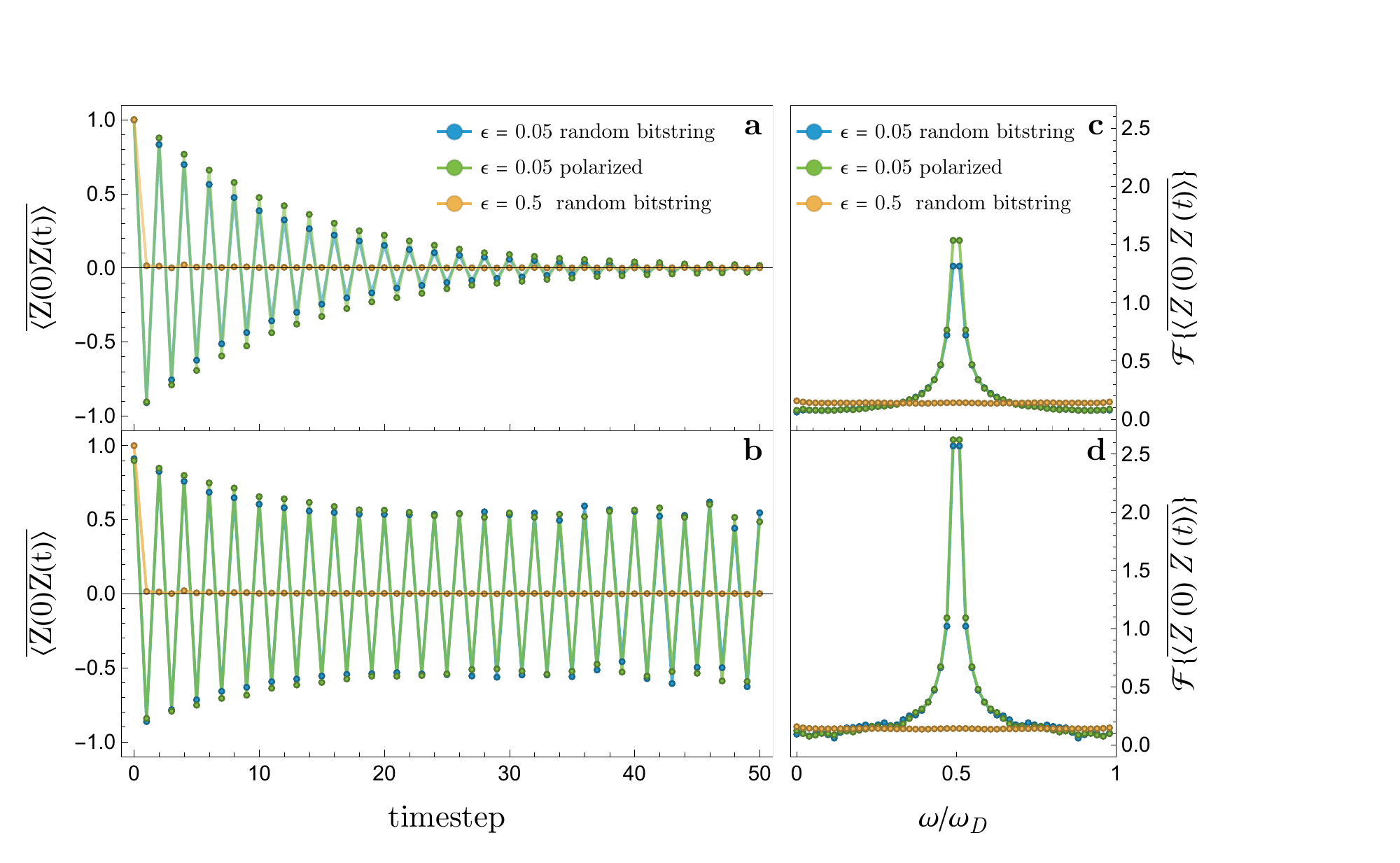}} \\
\subfloat{\includegraphics[scale=0.6]{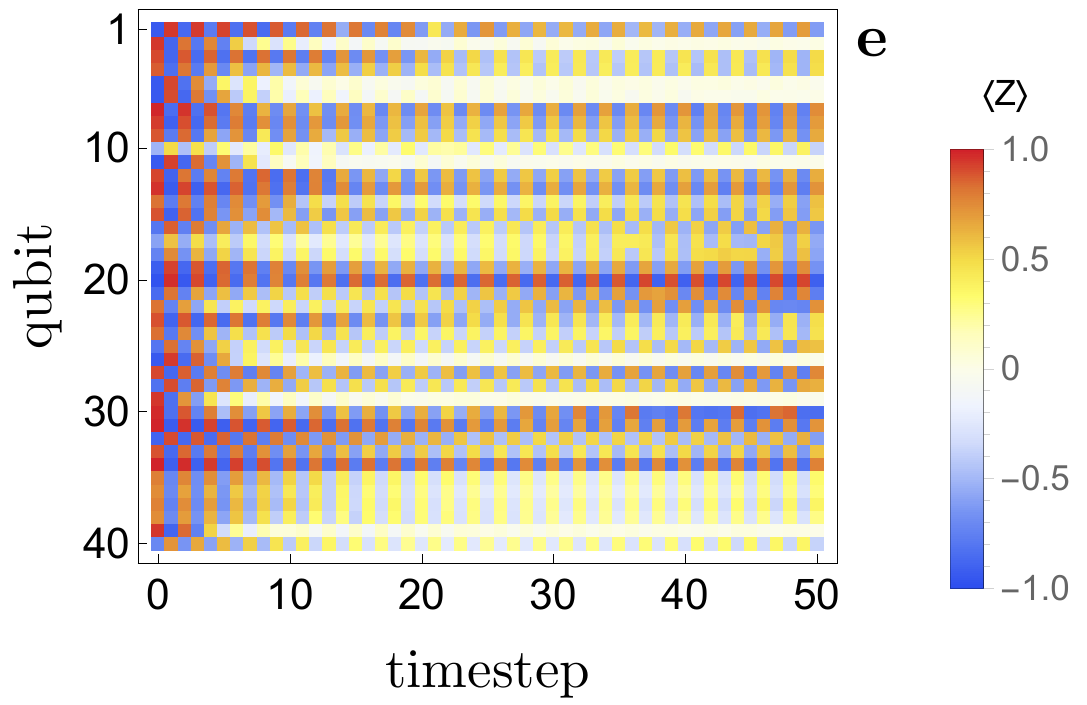}  \includegraphics[scale=0.6]{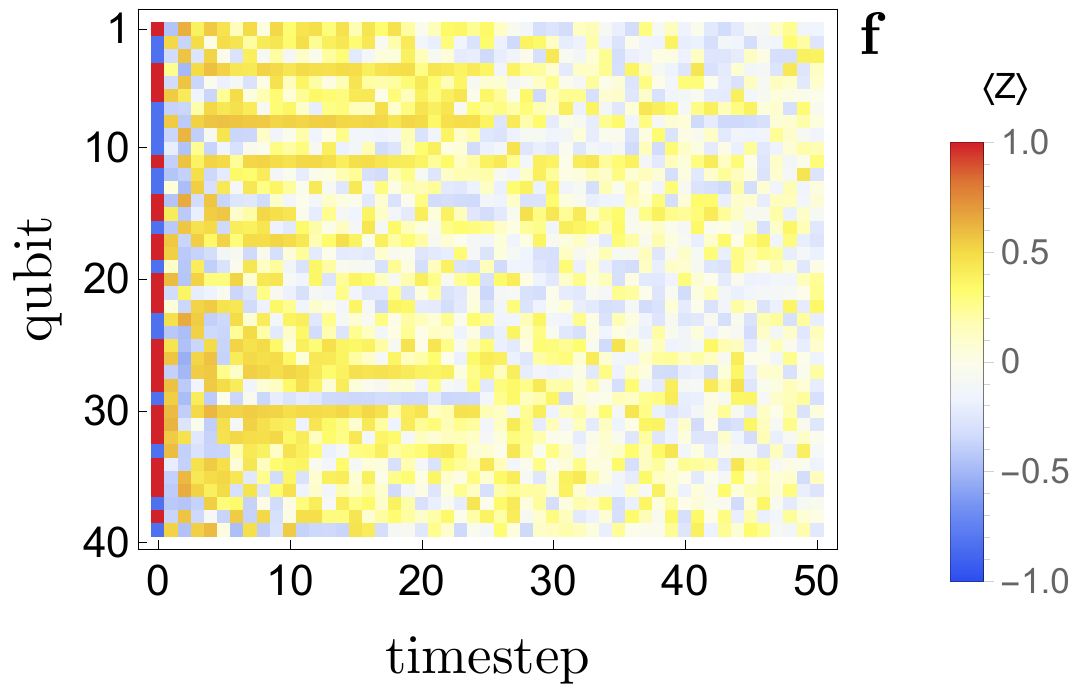}}
\caption{\textbf{Signatures of DTC and thermal dynamics}. \textbf{a} Averaged spin-spin auto-correlators across time with measurement error mitigation (see Methods) applied to the raw data. \textbf{b} Same but with additional correction of overall decay due to noise. \textbf{c} and \textbf{d} show corresponding frequency spectra. \textbf{e} and \textbf{f} show site resolved polarizations across time for $\epsilon = 0.05$ (DTC) and $\epsilon = 0.5$ (thermal phase), respectively. Of the total 57 qubits we show those 40 qubits that operate within reasonable bounds on error rates, as judged by our mitigation algorithm (see Methods).
}
\label{fig:RealTime}
\end{figure*}
%
\begin{figure*}[t!]
\centering
\includegraphics[width=0.99\textwidth]{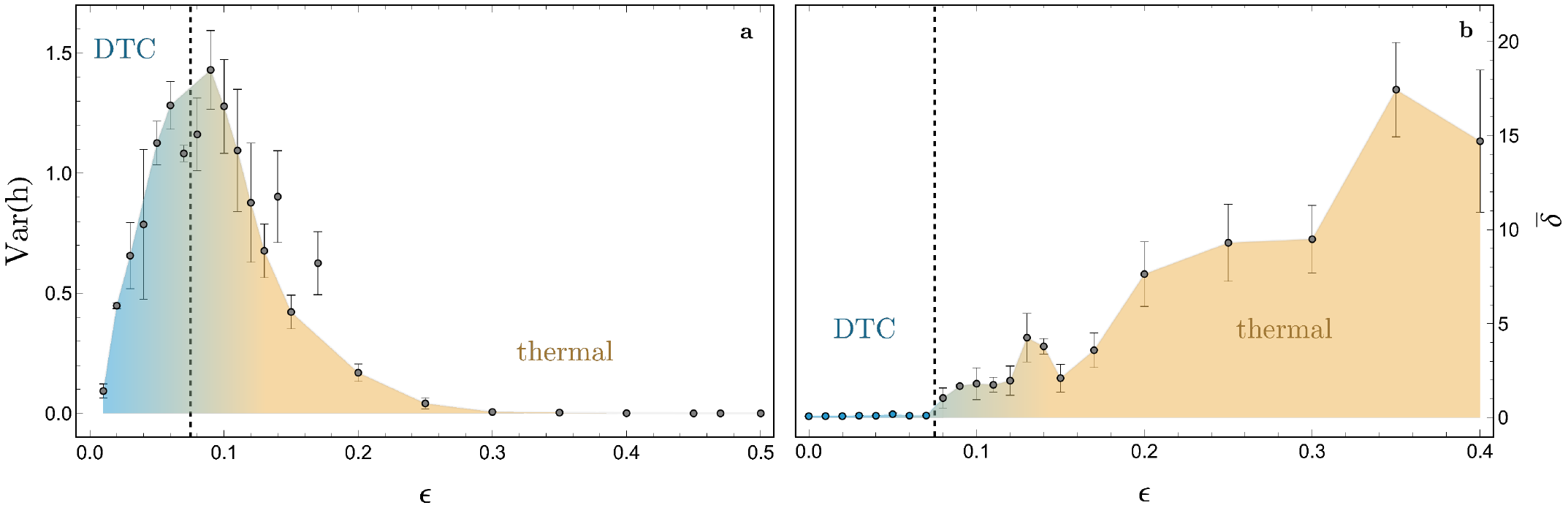}
\caption{\textbf{Dynamical phase transition}. \textbf{a} Critical fluctuations in the subharmonic frequency response. Filling and dotted line are a guide to the eye. \textbf{b} Average decay constant $\overline{\delta}$ across qubits. The transition from DTC to the thermal phase is indicated by the rather sudden increase in $\overline{\delta}$. The dashed vertical lines in a and b indicate the dynamical phase transition at $\epsilon_c\approx 0.075$.}
\label{fig:phasetransition}
\end{figure*}
%
The realization of the DTC can be observed by site-resolved measurement of the spin-spin auto-correlators across time, $\langle Z_i(0) Z_i(t)\rangle$. It is convenient to consider the averaged auto-correlator $\overline{\langle Z_i(0) Z_i(t)\rangle}~=~\sum_i \langle Z_i(0) Z_i(t)\rangle / N_q$ where $N_q$ is the number of qubits. Rapid decay of these correlators is expected for the thermal regime; in contrast, the oscillations in theory persist indefinitely for the DTC due to MBL. The inherent noise of NISQ devices causes depolarization of the qubits and represents the main challenge for distinguishing between DTC and thermal regime. Even in the ideal case of $\epsilon=0$, where $U$ should not generate any entanglement for an initial product state,
we observe a finite rate of decay. Fig.\,\ref{fig:RealTime}\,a shows the averaged autocorrelator after correcting for measurement errors. For $\epsilon=0.05$ we observe persistent oscillations as a hallmark of the realized DTC. Deviations from $\pm 1$ are attributed to a combination of noise and the fact that the conserved operators of the effective MBL Hamiltonian do not exactly coincide with the set of operators  $\{Z_i\}$, hence the $Z_i$ are only partly conserved at late times. Nonetheless, the oscillators are stable over 50 Floquet periods and are clearly distinct from prethermal dynamics\,\cite{kyprianidis-21s1192,ippoliti2021manybody}. Even with finite coherence time, one can still clearly distinguish between the rapid decay of a thermal system for $\epsilon=0.5$ and the DTC for $\epsilon=0.05$. In addition, we show the latter for a fully polarized initial state with almost identical results.
Using reference data obtained for $\epsilon=0$ allows us to rescale the auto-correlations, since in this case any deviation from a perfect oscillation between $-1$ and $1$ is caused by noise. The precise error mitigation scheme is more involved and has been detailed in the Method section. In Fig.\,\ref{fig:RealTime}\,b we show the fully error-mitigated data, corresponding to the data shown in panel a, including the correction for noise-induced depolarization. The data for $\epsilon=0.5$ leads to immediate thermalization; for $\epsilon=0.05$ we observe the persistent oscillations of a DTC, irrespective of which initial state we choose (see Supplement).

We note that the error mitigation scheme does not boost the observed signal beyond factoring out the baseline decay rate, as described in the Method section.

The DTC phase can also be observed through a pronounced peak in the Fourier spectrum of $\langle Z_i(0) Z_i(t)\rangle$
at half the driving frequency $\omega_\mathrm{D}=2\pi T^{-1}$ for each spin. We show the qubit-averaged Fourier spectrum in Fig.\,\ref{fig:RealTime}\,c~(d) that corresponds to the measurement-error-mitigated (fully error-mitigated) data. Even in the absence of any error mitigation, the peak at $\omega_\mathrm{D}/2$ is pronounced; the mitigation scheme almost doubles the peak height.

In Fig.\,\ref{fig:RealTime} e and f we show site resolved measurements for representative data points within the DTC phase and the thermal phase, respectively, for different initial states. {Displayed are the 40 qubits that fulfill the criteria for suffiently low error rates as defined in the Methods section. The former displays, to varying degree, staggered spin polarizations across time for each individual qubit, while the latter shows rapid depolarization across the entire system. This shows a clear distinction between two different dynamical phases. One phase is characterized by the breaking of ergodicity through MBL and furthermore the spontaneous breaking of an emergent Ising symmetry, resulting in period doubling and therefore time crystalline dynamics. The other phase exhibits standard ergodicity and thus rapidly evolves towards a thermal state.

\section*{Dynamical phase transition}

In the following, we focus on the dynamical phase transition from the DTC regime to the thermal phase.
It can be shown\,\cite{PhysRevLett.118.030401} that time evolution under the above Floquet unitary over an even integer number of periods is unitarily equivalent to time evolution with an effective Hamiltonian,
\begin{equation}\nn
    V \, U(2nT) \, V^{\dagger} \approx e^{-i 2n H_{\mathrm{TFIM}} T}  .
\end{equation}
The approximate sign indicates that the representation in terms of a conserved Hamiltonian is not correct out to temporal infinity and that the effective Hamiltonian itself contains higher order terms that we neglect\,\cite{else-17prx011026}.
$V$ is a finite depth unitary operator that depends on the particular disorder and the random transverse field Ising model (TFIM) Hamiltonian is given by
\begin{equation}\nn
    H_\mathrm{TFIM} = \sum_i \tilde{J}_i Z_{i} Z_{i+1} +  \tilde{B}_{i}^x \, X_{i}\ .
\end{equation}
The parameters
$\tilde{J}_{i}^{z}$ and $\tilde{B}_{i}^{x}$ are disordered and varying $\epsilon$ essentially translates into varying the mean $\overline{\tilde{B}_{i}^{x}}$. The random TFIM exhibits an Ising symmetry that its eigenstates either share or spontaneously break depending on the magnitude of $\overline{\tilde{B}_{i}^{x}}$. The latter case corresponds to the DTC phase of the Floquet system.

This leads us to the first indicator of the phase transition, namely critical fluctuations at the transition between the DTC and the thermal phase in the order parameter that spontaneously breaks the Ising symmetry in one phase but not in the other. The order parameter of the random Ising transition is the $z$-magnetization and a finite value will result in stable oscillations at $\omega = \omega_\mathrm{D}/2 $
due to the periodic flip operation. Defining $h_{i} = \abs{\mathcal{F}\{ \langle Z_{i}(t) \rangle \}(\omega_\mathrm{D}/2) }$, with $\mathcal{F}\{\cdot\}$ representing the Fourier transform, one can conclude that the variance across the chain $\mathrm{Var}(\{h_{i}\})$ should vanish in the DTC phase as well as in the thermal phase. At the transition we instead expect critical fluctuations to produce a finite value. For short chains ($ N \sim 10$) one has to average over several disorder realizations to obtain a clear signal since the behavior at the transition is sensitive to the particular choice of disorder. Our system of $N=57$ sites seems to produce a rather clear signal without this averaging process, as one might expect.
Fig.\,\ref{fig:phasetransition}\,a shows the variance as a function of $\epsilon$ after we have reduced the noise by applying error mitigation (see Methods). A pronounced peak indicates the phase transition and allows us to extract $\epsilon_c\approx 0.075$ (the peak's maximum) as an estimate for the phase transition. Finite values at very small values of $\epsilon$ may be attributed to the aforementioned fact that only part of $Z_i$ is actually conserved at late times.

The second indicator of the transition is the average decay constant $\overline{\delta}$ of the local polarization across the chain. With every spin roughly following an exponential decay (barring initial transients), i.e., $| \langle Z_i \rangle|(t) \propto \exp(-\delta_i t)$, we define $\overline{\delta} = \sum_i \delta_i / N_q $. While finite even in the ideal case of $\epsilon = 0$ due to noise, $\overline{\delta}$ is expected to increase significantly when transitioning into the thermal phase. The transition found in Fig.\,\ref{fig:phasetransition}\,b nicely agrees with $\epsilon_c\approx 0.075$.

\section*{Discussion}

The model as defined in Eqs.\,\ref{eq1} and \ref{eq2} is non-interacting. However, using process tomography we were able to analyze the systematic error contributions to the effective Hamiltonian associated with the Ising interaction. The dominant terms generated correspond to longitudinal fields and effectively contribute a third unitary $U_3=\exp{\left(-i\sum_i b_i Z_i\right)}$, where the random amplitudes $b_i$ can be at least as large as $\pi/25$. In addition, further spin-spin interactions are generated (see Supplement for details).
We therefore argue that the actual model as implemented on the quantum computer is truly interacting and thus qualifies as MBL. This is in line with the reasoning and numerical evidence put forward in Ref.\,\onlinecite{ippoliti2021manybody}.
As we do not explicitly add an additional longitudinal field unitary $U_3$ by hand, our circuit depths are signifantly reduced. This allows us to minimize the noise in our signal while utilizing the ubiquitous gate errors to meet one of the requirements for MBL.

It is well known that special states at the edge of the spectrum, such as a fully polarized chain or a Neel state, can exhibit prethermal dynamics that resembles DTC even if the full spectrum is not MBL. In that case one expects a strong dependence of the observed dynamics on the initial state. Random bit strings would show a significantly faster rate of depolarization than any of these special states. By comparing the data obtained for a fully polarized chain with the generic random bit string (see Fig.\,\ref{fig:RealTime}) we conclude that there seems to be no such strong dependence, especially when noise is taken into account. In the Supplement, we also show a fully polarized chain as well as a random bit string as initial states without any even-Ising disorder in the driving Hamiltonian, leading to a significant reduction in the oscillation amplitude as expected.

The DTC phase and the transition to the thermal phase could be observed for a significantly larger system size than in previous experiments and we were able to make a first step towards establishing a dynamical phase diagram for this system. With increasing fidelity of NISQ devices in the near-term future we can expect to shine light on some of the open problems in MBL, such as finite size scaling and the transition between the thermal phase and the MBL paramagnet.

\vspace{5pt}
{\it Note added:}~After being informed about Ref.\,\onlinecite{ippoliti2021manybody} we have updated the manuscript accordingly. During completion of this work we became aware of two related works. In Ref.\,\onlinecite{xu-21arXiv} a DTC phase using an array of eight capacitively coupled transmon qubits is demonstrated experimentally. In Ref.\,\onlinecite{google-21arXiv} the observation of a DTC on a quantum computer with 20 qubits was reported. Our main results in Fig.2 and Supplement Fig.5 are in agreement with Ref.\,\onlinecite{google-21arXiv}, but were derived independently.

\section*{Methods}

\noindent
\textbf{Error mitigation:}\\
For current NISQ devices error mitigation is crucial. Notably, there has been significant progress in developing error mitigation schemes and demonstrating stability of quantum circuits against quantum gate errors and shot noise\,\cite{PhysRevX.8.031027,PhysRevApplied.15.014023,Vallury2020quantumcomputed, vovrosh2021simple}. For this work, we have developed a scheme based on our understanding of the dominant contributions to the total error. Our error mitigation scheme is tailored to deal with measurement errors and the depolarization due to the environment.

\vspace{5pt}
\noindent
\textbf{Measurement error:}\\
The first major contribution is the error associated with measurement of the individual qubit's state. For a small set of $n$ qubits one can calibrate and correct for this error by initializing the register in each of the $2^n$ possible computational basis states and measuring $\langle Z \rangle$ on every qubit. However, for a large register this is not feasible and so we assume that the measurement errors on different qubits are approximately uncorrelated. Since the measurement errors are not consistent across different runs over the course of hours or days, one would have to recalibrate before every experiment and assume, that the drift is minimal in between calibration and the actual simulation. Systematic errors that offset the polarization might also be introduced by gate errors resulting from the application of trotterized time evolution. In order to tackle both of these effects we use an empirical approach. Along with every simulation run for a given, finite $\epsilon$ and fixed disorder and initial state we also perform a reference simulation with $\epsilon = 0$ and otherwise unchanged conditions. Any bit string is an eigenstate for the unperturbed unitary. This allows us to conclude that any deviation from a constant local polarization is due to noise and errors. The combination of measurement errors and systematic gate errors at late times is characterized by a pair of effective error parameters $\eta_0$ and $\eta_1$ for each qubit. $\eta_0$ denotes the effective probability of erroneously measuring a 1 given that the qubit is in state $\ket{0}$ and $\eta_1$ denotes the effective probability of obtaining a 0 from a $\ket{1}$ state. We are interested in the expectation value $\langle Z \rangle$ and with $\bar{\eta} := (\eta_0 + \eta_1)/2$
and $\Delta:=\eta_0 - \eta_1$, one can easily show that the corrected expectation value $\langle Z \rangle_\mathrm{corr}$ can be obtained from the measured one $\langle Z \rangle_\mathrm{meas}$ via
\begin{equation}
 \langle Z \rangle_\mathrm{corr} = \frac{\langle Z \rangle_\mathrm{meas} + \Delta}{1-2\bar{\eta}}\ .
\end{equation}
This empirical approach requires the input of at least two data points at different times. In order to not overcorrect the measurements at early times, where systematic gate errors have not yet accumulated to the same extend as for late times, we choose to normalize the polarisiations to $|\langle Z \rangle(t=0)| = 1$:
\begin{equation}
 \langle Z \rangle_\mathrm{corr} = \frac{\langle Z \rangle_\mathrm{meas} - \langle Z \rangle_\mathrm{final}}{| \langle Z \rangle_{meas}(t=0)- \langle Z \rangle_\mathrm{final} |}\ .
\end{equation}
$\langle Z \rangle_\mathrm{final}$ represents an average over a few data points at the latest simulated times. The average is taken in order to avoid the effects of changing conditions over the course of the simulation. From this one can in principle extract the effective empirical error paramters:
\begin{align}
 \eta_0 &= (1 - | \langle Z \rangle_\mathrm{meas}(t=0)- \langle Z \rangle_\mathrm{final} | - \langle Z \rangle_\mathrm{final})/2\ ,     \\[5pt]
 \eta_1 &= (1 - | \langle Z \rangle_\mathrm{meas}(t=0)- \langle Z \rangle_\mathrm{final} | + \langle Z \rangle_\mathrm{final})/2\  .
\end{align}

\vspace{5pt}
\noindent
\textbf{Depolarization due to environment:}\\
In addition to the measurement and gate errors one can also observe an overall exponential decay of polarization due to the qubits thermalizing with their environment. To compensate for this decay, we would like to rescale the data accordingly. We use the following algorithm in order to distinguish between a qubit being non-polarized at late times due to internal thermalization versus due to erroneous spin flips:

\vspace{5pt}

\textbf{1.}~Introduce a cutoff parameter $W_0 \in [0,1]$, where 1 corresponds to the maximal polarisation of a qubit. Qubits whose average polarization over 5 timesteps drops below $W_0$ after the initial 13 timesteps for $\epsilon=0$ are excluded from the evaluation (further justification below).

\vspace{5pt}

\textbf{2.}~Introduce a fixed threshold $W_\mathrm{f} < W_0$ for finite values of $\epsilon$. Qubits that lie within the interval $[-W_\mathrm{f}, W_\mathrm{f}]$ after the initial time evolution are assumed to approach vanishing magnetization due to thermalization, in which case the accumulated gate error in the magnetisation is less significant ($\langle \sigma^z \rangle = 0 $ is a fixed point as far as the Trotter gate errors are concerned). Therefore we do not rescale the subsequent data points for these qubits.

\vspace{5pt}

\textbf{3.}~For qubits outside of this interval one can assume that most of the damping is due to the same depolarising noise that was observed previously for $\epsilon=0$. We rescale the subsequent data points with an exponential fit obtained from the $\epsilon=0$ data for this qubit but adjusted by the ratio of average polarisation for given, finite $\epsilon$ to the corresponding average polarisation for vanishing $\epsilon$. With $m^{(\epsilon)}_{i}(t)$ denoting the magnetisation of the $i$th qubit at time $t$ as measured on the QC and corrected for measurement errors, the exponential fit of the reference data at $\epsilon=0$ takes the form
    \begin{equation}
        \frac{1}{2}\Big[m^{(0)}_{i}(t)+\mathrm{sign}\left(m^{(0)}_{i}(t)\right)\Big] = a^{(0)}_{i} e^{- b^{(0)}_{i} t} + c^{(0)}_{i}
    \end{equation}

    with fit parameters $a^{(0)}_{i}$, $b^{(0)}_{i}$, and $c^{(0)}_{i}$.
    \noindent We define the rescaled magnetisation $M^{(B)}_{i}(t)$ via

    \begin{equation}
        M^{(\epsilon)}_{i}(t) = \frac{ \overline{m^{(\epsilon)}_{i}} }{ \overline{m^{(0)}_{i}} } \frac{ m^{(\epsilon)}_{i}(t)+\mathrm{sign}\left(m^{(\epsilon)}_{i}(t)\right) }{ a^{(0)}_{i} e^{- b^{(0)}_{i} t} + c^{(0)}_{i}} - \mathrm{sign}\left(m^{(\epsilon)}_{i}(t)\right) .
    \end{equation}

    The bar denotes an average over 5 timesteps after the initial 13 timesteps. Omitting the first few iterations serves to avoid the influence of initial transients on the observed signature.

\begin{figure}[t!]
\centering
\includegraphics[clip, trim=1.5cm 2cm 1.5cm 2cm ,width=0.7\columnwidth]{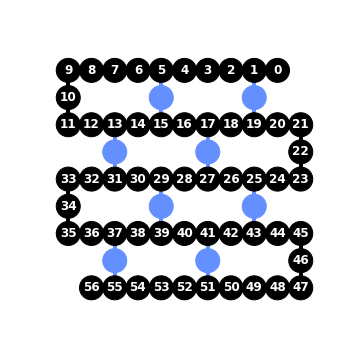}
\caption{Qubit layout on {{\it ibmq\_manhattan}} and {{\it ibmq\_brooklyn}} chips with its 65 qubits (for the former whole-device entanglement was demonstrated recently\,\cite{mooney-21arXiv2102.11521}). The 57 black qubits are used for the simulations of DTC.}
\label{fig:layout}
\end{figure}

Based on numerical simulations for smaller chains one expects the individual spins to approach an almost steady state after intial, short lived transients. One therefore predicts that most spins would exhibit an exponential decay even for finite $\epsilon$. Since not all of the 57 qubits seem to operate within the margins given by the stated error rates, we additionally remove those few ones that deviate significantly from the exponential decay model as judged by convergence of our fit. The above procedure requires us to omit qubits whose error rates are sufficiently high to move their absolute value polarization into the interval $[0,W_0]$ even for $\epsilon=0$, because we would otherwise confuse them for ones that are thermal. This justifies step 1.\ in the above error mitigation scheme. \\
The same error mitigation algorithm was applied to all of the data in Fig.\,\ref{fig:RealTime}\,b. Rescaling of the late-time data points only occurs if every one of the above mentioned criteria is met. This ensures that instances of significantly dampened time  evolution, as displayed by the yellow curve in Fig.\,\ref{fig:RealTime} remain dominated by the increased rate of depolarization.  \\

In order to ensure that the arbitrary threshold $W_0$ does not significantly affect the results we process the data using a range of different values. The peak associated with critical fluctuations and the signature of increased damping remain quite stable across a wide range. The same applies to the choice of  $W_\mathrm{f}$ that enters in the definition of the interval. Here values very close to 1 produce an algorithm that is too sensitive to fluctuations in the error rates, while values close to 0 suppress the signal. The results presented in this paper were evaluated using $W_0 = 0.15$ and $W_\mathrm{f}/W_0 = 2/3$. We find that typically $\sim\,40-45$ qubits enter the evaluation.

\section*{Data availability}
Data that support the findings of this paper are available from the corresponding author upon reasonable request. 

\section*{Code availability}
The code of the quantum circuits of the study are available from the corresponding author upon reasonable request.

\section*{Acknowledgements}
The authors acknowledge valuable discussions with Vedika Khemani, Roderich Moessner, Gregory White, Charles Hill, Lloyd Hollenberg, Thomas Quella and Jan de Gier.
This work was supported by the University of Melbourne through the establishment of an IBM Quantum Hub at the University.
SR acknowledges support from the Australian Research Council through Grant No.\ FT180100211.

\section*{Competing interests}
The authors declare no competing interests.

\bibliography{Bibliography.bib}

\end{document}


\renewcommand{\vec}[1]{\boldsymbol{#1}}
\newcommand{\up}{{\uparrow}}
\newcommand{\dw}{{\downarrow}}
\newcommand{\pa}{{\partial}}
\newcommand{\pd}{{\phantom{\dagger}}}
\newcommand{\bs}[1]{\boldsymbol{#1}}
\newcommand{\todo}[1]{{\textbf{\color{red}ToDo: #1}}}
\newcommand{\new}[1]{{\textbf{\color{red}#1}}}
\newcommand{\sr}[1]{{\color{blue}#1}}
\newcommand{\srr}[1]{{\color{orange}#1}}
\newcommand{\eps}{{\varepsilon}}
\newcommand{\I}{{i\mkern3mu}}
\newcommand{\nn}{\nonumber}
\newcommand{\ie}{{\it i.e.},\ }
\def\eg{\emph{e.g.}\ }
\def\ea{\emph{et al.}}
\def\cf{\emph{c.f.}\ }


\title{Supplement:\\
Realization of a discrete time crystal on 57 qubits of a quantum computer}

\author{Philipp Frey}
\author{Stephan Rachel}
\affiliation{School of Physics, University of Melbourne, Parkville, VIC 3010, Australia}

\date{\today}


\maketitle

\section*{S1.\ Comparison of different initial states and absence of disorder}

Here we show the results for a Neel state, in addition to Fig.\,2 in the main text. Moreover, we study both a fully polarized and a random-bit state for vanishing Ising-even disorder in the driving Hamiltonian. The decay of the Neel state matches that of a polarized or random initial state very closely in the presence of disorder, indicating MBL. Without disorder the random initial state depolarizes significantly faster than any of these. This becomes even more apparent after applying the error mitigation where the inconsistent oscillations at late times typically indicate that the unmitigated signal is too weak and the result of the mitigation scheme become error dominated. A fully polarized chain without disorder in the driving Hamiltonian seems to decay signifantly faster compared to the disordered system as well. The difference becomes more apparent after the error mitigation. However, it seems to be more stable than the random bit string, indicating that special initial states can exhibit prethermal dynamics that mimics DTC even in the absence of MBL.

\begin{figure*}[h!]
\centering
\hspace{13mm}
\subfloat{\includegraphics[width=0.9\linewidth]{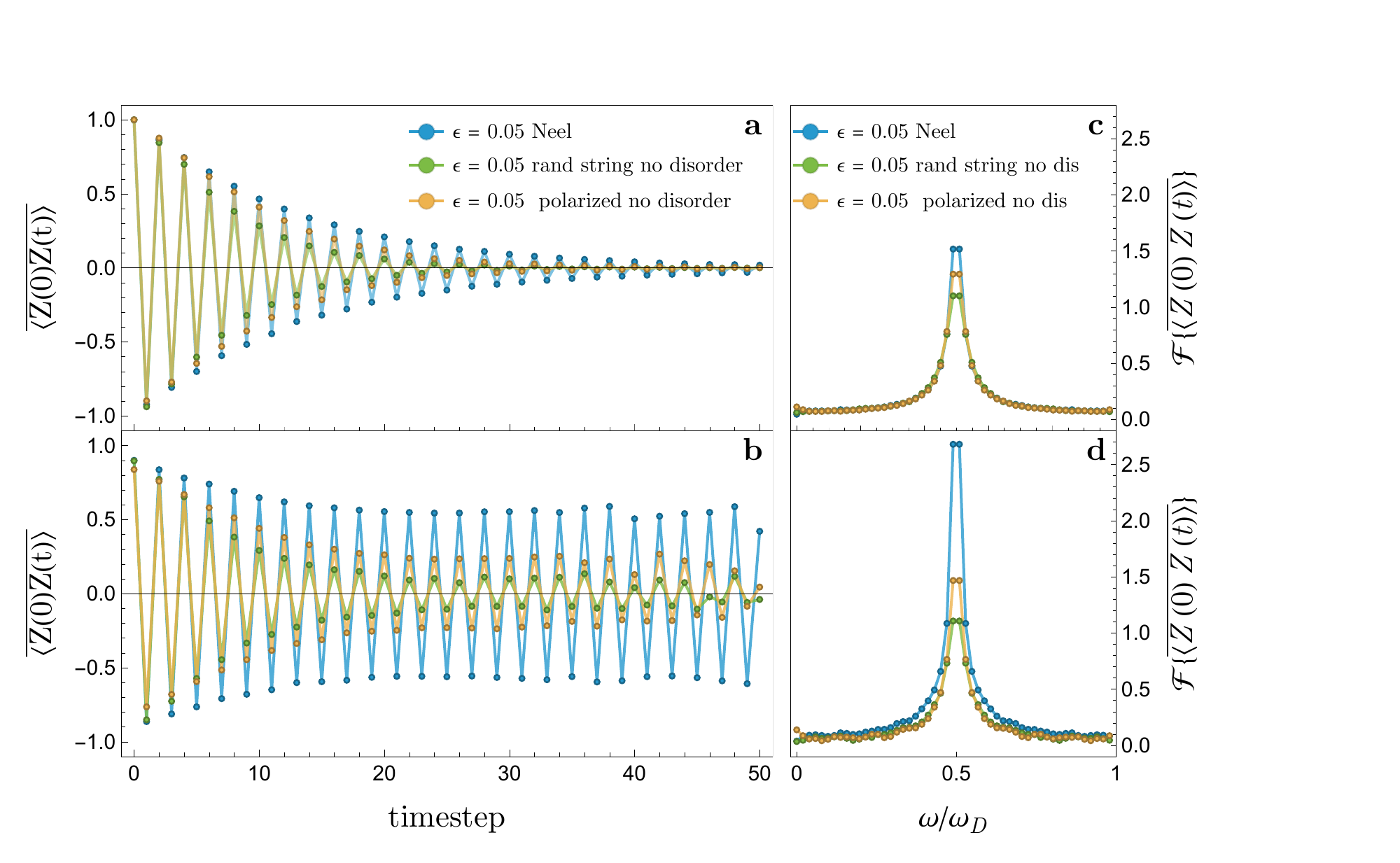}}
\caption{\textbf{DTC stability of special states and translation invariant chains.} \textbf{a} Averaged spin-spin auto-correlators across time with measurement error mitigation (see Methods) applied to the raw data. \textbf{b} same but with additional correction of overall decay due to noise. \textbf{c} and \textbf{d} show corresponding frequency spectra.}
\label{fig:RealTimeSup}
\end{figure*}

\newpage
\section*{S2.\ Local spin observables}

\begin{figure}[h!]
\centering
\includegraphics[width=0.4\columnwidth]{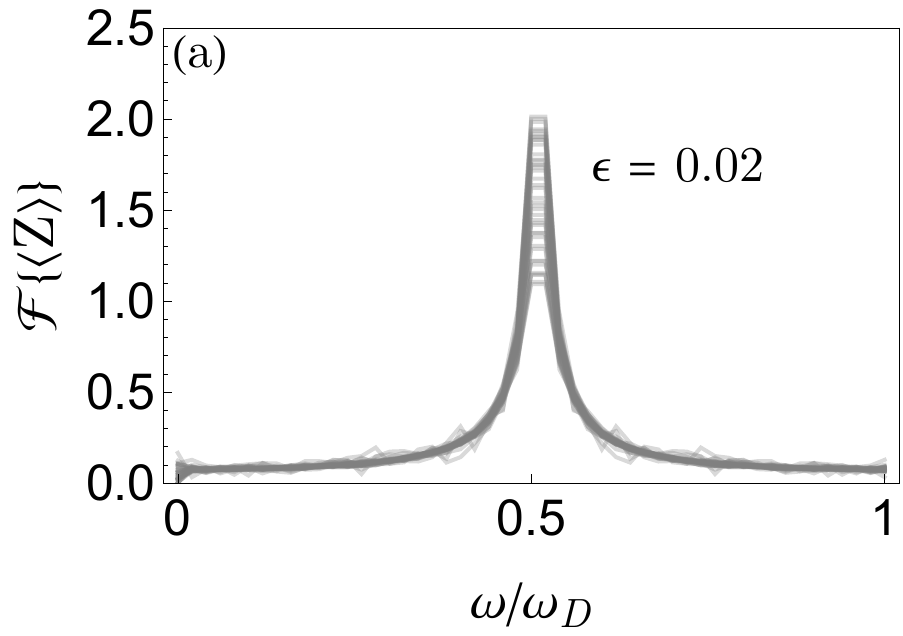}
\includegraphics[width=0.4\columnwidth]{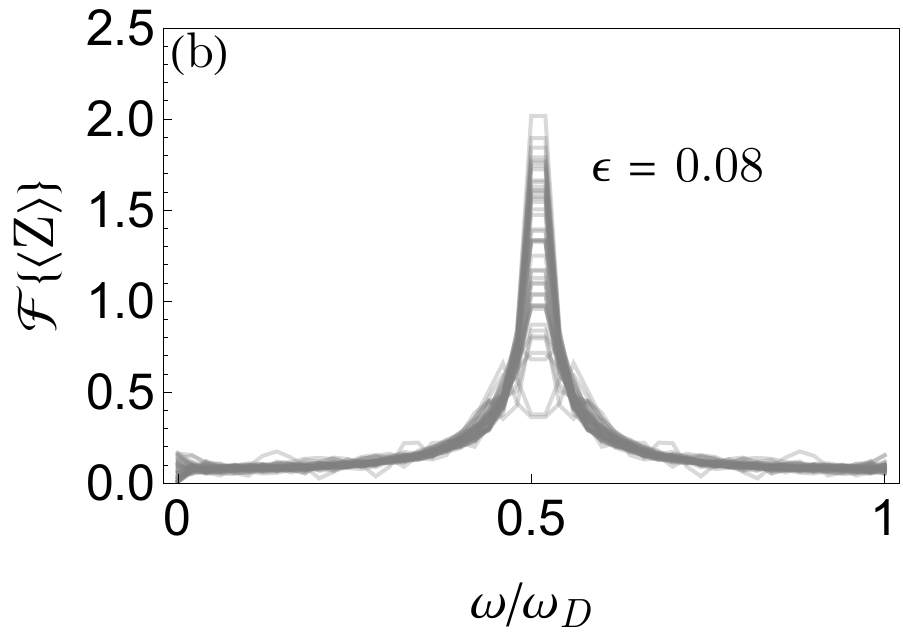}
\includegraphics[width=0.4\columnwidth]{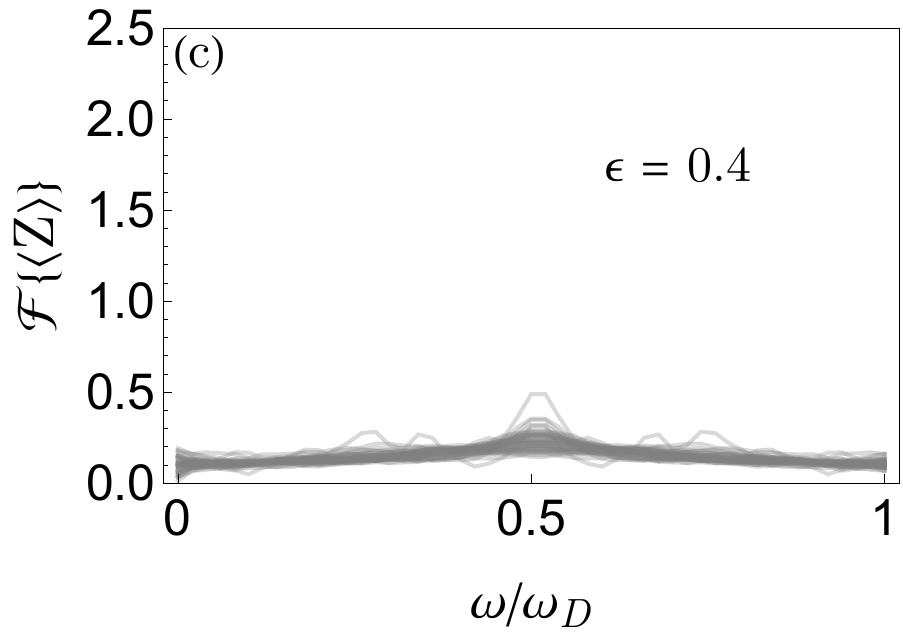}
\caption{\textbf{Site resolved spin observables} \textbf{a} Overlay of Fourier spectra of each qubit for $\epsilon=0.02$. \textbf{b} and \textbf{c} show the same for $\epsilon=0.08$ and $\epsilon=0.4$ respectively.}
\label{fig:subharmResp}
\end{figure}

Here we show the site resolved Fourier spectrum (absolute value) of oscillations in $\langle Z(t) \rangle$ for different values of $\epsilon$. This equivalent to looking at $\langle Z(0) Z(t) \rangle$ since the difference in sign does not affect the absolute values of the individual Fourier spectra. The pronounced peak at half the driving frequency $\omega_D$ at low $\epsilon$ indicate the DTC phase. Fig.\ref{fig:subharmResp}\,a with $\epsilon=0.02$ is supposedly well inside the DTC phase but clearly shows that there is a finite variance in the amplitude of the subharmonic frequency response. This is due to the fact that only a finite fraction of each spins $Z$-component is conserved and this fraction varies accross the chain due to the quenched disorder. Well within the thermal phase none of the spins are oscillating at late times and hence the frequency response is almost completely flat, see Fig.\ref{fig:subharmResp}\,c. Thus the associated variance vanishes as well. In the vicinity of the phase transition the excess variance compared to the case of small $\epsilon$ becomes very apparent in the site resolved spectrum, Fig.\ref{fig:subharmResp}\,b.

\newpage
\section*{S3.\ Site resolved DTC across entire chain}

Here we show the site resolved oscillations after applying full error mitigation but without removing any qubits based on the filter criteria presented in the Methods section. It is clear that the increasing signal towards late times on some qubits (here qubits 20, 21, 22, 24, 30) is an artefact of the mitigation scheme and typically avoided when filtering out qubits based on particularly high error rates. It also seems that many of the qubits that thermalize rapidly and completely (here qubits 11, 25, 27, 42, 45) are typically removed by our scheme, indicating that this is due to significantly higher error rates.

\begin{figure}[h!]
\centering
\includegraphics[width=0.6\columnwidth]{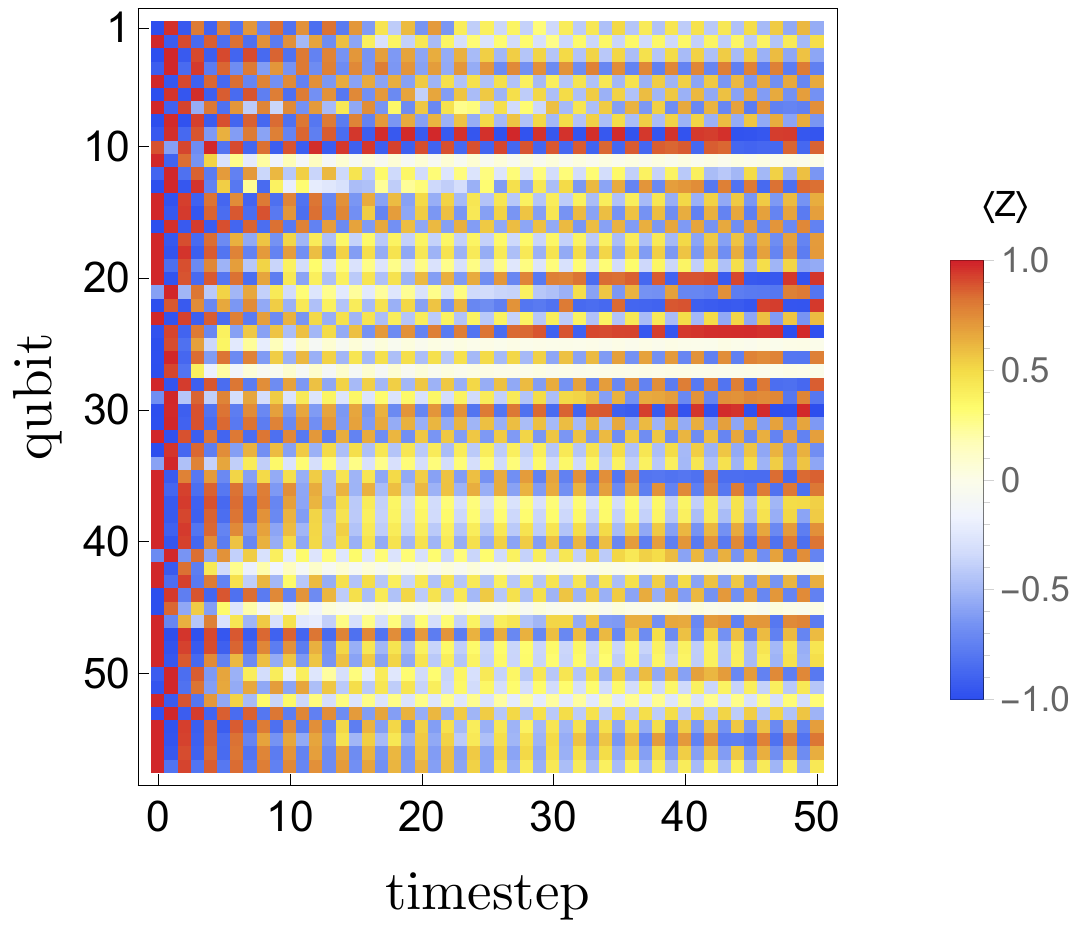}
\caption{\textbf{Site resolved DTC including all qubits.} Fully error mitigated data obtained for $\epsilon = 0.05$ on {\it imb\_brooklyn}.}
\label{fig:MatrixPlotDTCAllQubits}
\end{figure}

\newpage
\section*{S4.\ Process Tomography}

In the following we present the results for process tomography performed on a single Floquet time step on three qubits. Process tomography consists of preparing the quantum register in a variety of different initial states, executing the quantum circuit in question, and then measuring the output state with respect to different bases. From the shot statistics one extracts an estimate of the quantum map associated with this process, which may differ from the ideal unitary due to error contributions.

In particular one can define the post-gate error generator for the map, $L = \log(G*H^{-1})$, where $G$ is the process matrix for the estimate and $H$ is the process matrix for the ideal gate. This describes all Markovian errors as if they occurred after the gate. The error generator decomposes like a Lindbladian into coherent Hamiltonian terms and dissipative terms. The Hamiltonian generators are the relevant part for us, since they constitute additional effective terms in the Floquet unitary that are systematic and not included in the original model. Since the data for each time step in our simulation is averaged over many shots, it is affected by exactly these systematic terms.


Table\,\ref{tab:processtomography} shows the coefficients of the dominant Pauli terms in the effective ``post-gate Hamiltonian" $H_\mathrm{add}$, i.e., our unitary $U$ gets modified to $\exp(- \I H_\mathrm{add}) * U$. In particular, we observe the local and random $Z$ rotations, mentioned in the main text, as well as significant weight-two interactions (i.e., terms such as $Z_i X_j$), thus substantiating our claim that the effective model is interacting.
To be specific, $H_{\rm add}$ contains all the terms listed in Table\,\ref{tab:processtomography}, i.e., $H_{\rm add} = 0.118 X_3 + 0.085 Y_3  + 0.126 Z_3 + 0.023 X_2 + \ldots$, where we dropped the remaining nine terms.

\begin{center}
\begin{table}[h!]
\begin{tabular}{||c | c||}
 \hline
 ~Pauli operator~ & ~coefficient~  \\ [0.5ex]
 \hline\hline
 IIX & 0.118  \\
 \hline
 IIY & 0.085  \\
 \hline
 IIZ & 0.126  \\
 \hline
 IXI & 0.023  \\
 \hline
 IYI & 0.012  \\
 \hline
 IZI & 0.033  \\
 \hline
 IZX & 0.038  \\
 \hline
 IZY & 0.033  \\
 \hline
 XII & 0.024  \\
 \hline
 IXI & 0.023  \\
 \hline
 YZI & 0.023  \\
 \hline
 ZII & 0.037  \\
 \hline
 ZYI & 0.017  \\ [0ex]
 \hline
\end{tabular}
\caption{\textbf{Dominant Pauli operators contributing to the effective additional unitary.} Terms and coefficients were derived via process tomography on three qubits. For instance, ``IIX'' is a short notation for the term $I_1 I_2 X_3$. Besides non-trivial spin-spin interactions, longitudinal fields (IIZ, IZI, ZII) are present with amplitudes 0.126, 0.033 and 0.037, respectively.}
\label{tab:processtomography}
\end{table}
\end{center}

We emphasize that extracting the Pauli operators and their amplitudes from the quantum computer data is performed on a classical computer, preventing us from extracting the same information for all 57 qubits. This is, however, not necessary: the results of our process tomography on three qubits are expected and appear to be consistent with previous analysis of IBM's qubits\,\cite{malekakhlagh-20pra042605,govia-20nc1084}. Our findings outlined in Table\,\ref{tab:processtomography} are thus representative for the 57 qubit circuits in the main text.


\bibliography{Bibliography.bib}